\documentclass{article}
\usepackage{spconf,amsmath,graphicx,amsfonts, hyperref}
\usepackage{xcolor}

\title{Enhancing Affective Representations of Music-Induced EEG through Multimodal Supervision and latent Domain Adaptation}

\name{Kleanthis Avramidis$^{1,2}$\quad Christos Garoufis$^2$\quad Athanasia Zlatintsi$^2$\quad Petros Maragos$^2$}
\address{\normalsize $^1$ Signal Analysis and Interpretation Lab, University of Southern California, Los Angeles, CA 90089, USA \\ \normalsize $^2$ School of ECE, National Technical University of Athens, 15773 Athens, Greece \vspace{-0.3cm}
}

\begin{document}
\ninept
\maketitle
\footnotetext[2]{This research work was performed at NTUA.}
\begin{abstract}

The study of Music Cognition and neural responses to music has been invaluable in understanding human emotions. Brain signals, though, manifest a highly complex structure that makes processing and retrieving meaningful features challenging, particularly of abstract constructs like affect. Moreover, the performance of learning models is undermined by the limited amount of available neuronal data and their severe inter-subject variability. In this paper we extract efficient, personalized affective representations from EEG signals during music listening. To this end, we employ music signals as a supervisory modality to EEG, aiming to project their semantic correspondence onto a common representation space. We utilize a bi-modal framework by combining an LSTM-based attention model to process EEG and a pre-trained model for music tagging, along with a reverse domain discriminator to align the distributions of the two modalities, further constraining the learning process with emotion tags. The resulting framework can be utilized for emotion recognition both directly, by performing supervised predictions from either modality, and indirectly, by providing relevant music samples to EEG input queries. The experimental findings show the potential of enhancing neuronal data through stimulus information for recognition purposes and yield insights into the distribution and temporal variance of music-induced affective features.
\end{abstract}

\begin{keywords}
Music Cognition, Emotion Recognition, Electroencephalography, Cross-Modal Learning \vspace{-0.14cm}
\end{keywords}

\section{Introduction}
\label{sec:intro}
\vspace{-0.15cm}
Music is an abstract, yet densely emotional form of art. It is universally enjoyed, due to its ability to induce powerful emotions irrespective of the underlying mood \cite{jakubowski2021} and has been characterized to greatly affect the function of the human brain~\cite{sacks,koelsch}. Hence it is widely used to study emotion recognition, both by analyzing the mood produced by several musical features \cite{panda,Song2012} and by studying its effects on human neural and physiological responses \cite{greer2019}. However, the task of extracting emotional information from brain activity poses severe challenges, due to the inherently abstract nature of the induced emotions, the variability in emotional and physiological responses between different individuals and the lack of large-scale databases of emotionally coordinated neural activity. In this paper, we propose a deep multimodal approach \cite{multimodaldeep}, using musical stimuli.

The scope of this study lies at the intersection of Music Cognition, Emotion Recognition from neuronal signals and Multimodal Learning. We choose to study brain responses to music by employing a cross-modal system to identify the correspondence between these modalities. We use the electroencephalogram (EEG) to model brain responses for this task and we constrain the learning process with emotion labels. Therefore, we aim to derive important insights regarding the affective role that music can play on humans and the extent to which it can help us build richer neuronal representations of affect. To conduct the experiment, we exploit multimodal optimization and domain adaptation strategies to project EEG and music features onto a common latent space, from which we could assess their similarity. By conditioning the learning process with emotion tags, the constructed space represents affect, enabling thus emotion recognition both directly, by performing supervised predictions, and indirectly, by ranking music tracks to EEG inputs, based on their distance. To the best of our knowledge, this is the first study to propose such a framework, thus it could be utilized as a baseline reference. We also perform an extensive qualitative study across 32 subjects of the DEAP dataset \cite{koelstra} to derive inter-subject affective patterns.

The remainder of this paper is organized as follows: Section 2 reviews the related work in cross-modal learning and research on EEG processing and music cognition. In Section 3 we introduce our problem and present the proposed framework along with the optimization methods we utilized. Section 4 includes information about the data, their pre-processing and implementation details, while in Section 5 we provide the experimental results. In that Section and in Section 6 we provide extensive quantitative and qualitative analysis of the outcomes, while Section 7 concludes our study. 
\vspace{-0.25cm}

\section{Related Work} \vspace{-0.15cm}
\label{sec:related}

\textbf{Music Cognition}: Studying the human brain's responses to music stimuli has always been a lively field of research in neuroscience and signal processing \cite{why} aiming to answer fundamental questions regarding our enjoyment of music. The field has gained a lot of attention in recent years, with the upsurge in available neuronal data. Many studies in the field rely on EEG recordings as they provide better temporal resolution than other techniques, such as functional magnetic resonance imaging (fMRI). In addition to the traditional, well-controlled auditory experiments, modern approaches gather physiological data from music listeners as they enjoy or imagine naturalistic music \cite{nmed}, in order for instance to examine correlations in temporal structure \cite{mind_the_beat} or the perceived tempo \cite{stober_tempo}. One of the core findings on music cognition is the correlation between characteristics of the neural oscillation patterns and rhythmical patterns in music \cite{Nozaradan10234}. Additionally, Event-Related Potentials (ERP) have been utilized to extract brain activity patterns that can relate to note onsets or pitch \cite{schafer,poikonen}. In parallel to the above, there has also been a shift towards deep learning approaches for information retrieval from music stimuli \cite{deep}, in which we focus on the present study.

\begin{figure*}
 \centerline{
 \includegraphics[scale=0.38]{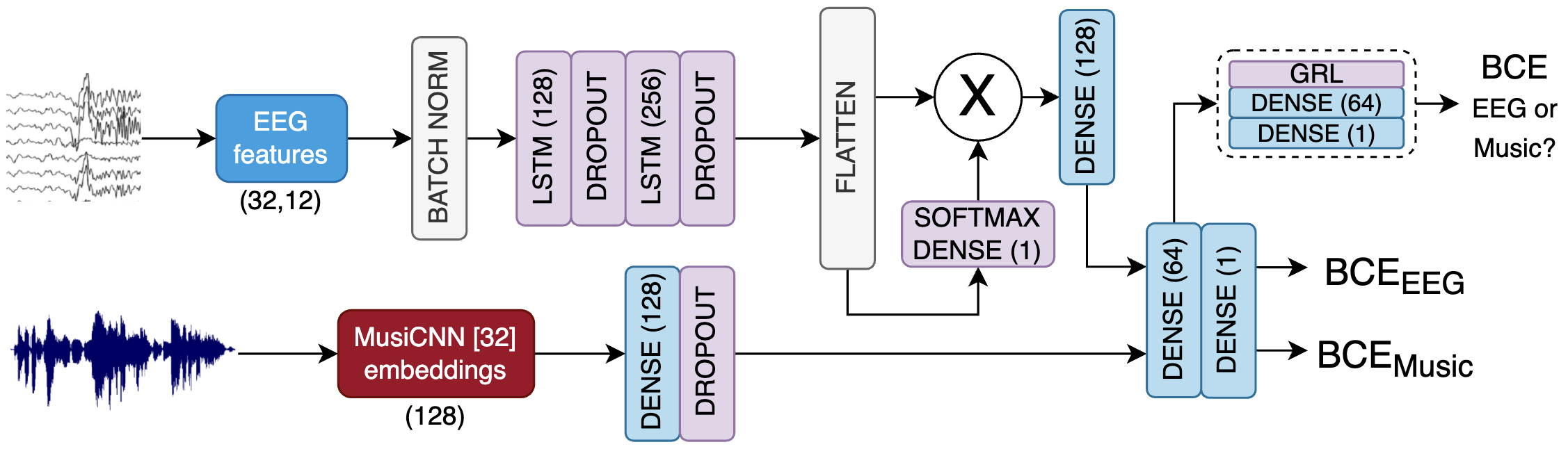}} \vspace{-0.3cm}
 \caption{The proposed bi-stream network. The output embedding layer of each stream is fed to the common 64D dense layer (common space).}
 \label{fig:bistream}
 \vspace{-0.3cm}
\end{figure*}

\textbf{Emotion Recognition}: Undoubtedly, the most powerful impact of music on humans concerns the induced emotions. Emotion Recognition is a widely researched field of contemporary Machine Learning and Behavioral Signal Processing \cite{em_survey} and several studies have focused on the musical features \cite{Song2012,panda2015} that determine affective attributes of music listening in a wide range of emotions.

Recently, several studies have examined physiological signals to analyze humans' felt emotions~\cite{phys}, with music emerging as an efficient method to elicit them. Due to its temporal resolution, the electroencephalogram is the most widely researched signal of this type and various statistical, spectral or time-frequency features have been proposed for Emotion Recognition \cite{spectral, hog}. Due to the noisy structure of EEG signals, many studies incorporate entropy \cite{dasm} and fractal \cite{fractals} algorithms to extract emotion-related features. Of course, variations of deep neural networks have been proposed and exceeded the performance of traditional feature extraction methods \cite{3dcnn, du_affcomp}, however the limited data availability and inter-subject variability present serious barriers for this kind of modeling.

\textbf{Cross-Modal Learning}: The task of learning a shared embedding space from different datasets or modalities is being studied through various approaches, which are predominantly applied to image and text modalities. A widely used baseline is Canonical Correlation Analysis (CCA). CCA is non-probabilistic and enables the extraction of linear components to optimize the correlation of pairs of vectors. One can find in the literature various non-linear CCA-based frameworks and architectures utilized to learn inter-modal similarities, such as Deep CCA \cite{dcca}. Besides CCA, other methods that have been used include an HGR-based maximal correlation metric \cite{hgr} and adversarial training \cite{acmr}, focusing mainly on the optimization function of the respective model, and on adaptive hidden layers~\cite{hu2018}. Another study \cite{LiK19} incorporated music to co-train a shared space with images using a contrastive loss. Further, in \cite{dscmr} a state-of-the-art framework exploits label supervision to better manipulate the latent space, a key concept that we also follow in our study.

\section{Methodology}
\label{sec:method}

In this study, we extract the semantic relationship between music tracks and corresponding EEG recordings, so that an EEG could be mapped to an efficient affective representation and retrieve emotionally consistent music samples. Let us assume a collection of $n$ instances of EEG-music pairs, denoted as $T = \{(x_i^a,x_i^b)\}_{i=1}^n$ where $x_i^a$ is the input EEG sample of the i$^{th}$ instance and $x_i^b$ the input music stimulus corresponding to that sample. Each instance has been assigned an affective annotation $y_i \in \mathbb{R}^2$ for valence and arousal dimensions. For each instance $i$ we aim to learn an EEG embedding $u_i = f(x_i^a,Y^a) \in \mathbb{R}^d$ and a musical audio embedding $v_i = g(x_i^b,Y^b)  \in \mathbb{R}^d$, where $d$ is the dimensionality of the common representation space and $Y^a, Y^b$ the trainable parameters.

\subsection{The Proposed Framework}

We use a bi-stream Neural Network with one branch for each modality. The EEG branch is a recurrent network, comprised of two LSTM modules and a softmax attention layer. The model takes as input an EEG trial of shape (channels, features) and attempts to capture its inter-channel correlations. Next, the output features of all channels are flattened and passed through an attention module to identify the most important components, that will lead the embedding vector in the common space. We utilize a lightweight network in order to avoid overfitting to the limited range of the available data, however any state-of-the-art model in the task could be applied. For the music branch we use the MusiCNN model \cite{musicnn} to extract high-level embeddings from the available audio stimuli. MusiCNN is a robust network, pre-trained on large audio databases, and produces high-quality music embeddings that compensate for the limited size of our track set and further assist the learning process. The extracted embeddings are then fed into a feed-forward neural network.

To construct the final bi-modal framework, we connect the last layer of each of the previous networks to a dense layer (Fig.~\ref{fig:bistream}) constituting the common representation space, from which we output emotion predictions. Inspired by \cite{du_affcomp}, we further apply a Gradient Reversal Layer (GRL) \cite{grl}, aiming to reduce the distribution shift between EEG and music modalities. In specific, both 64D embeddings are fed to this layer, from where we output a prediction regarding the modality type. From each batch, we randomly permute half of the EEG samples and their respective music samples, forming a new equal-sized batch that we shuffle and input to the GRL module, along with a binary label vector to denote the modality. Subsequently, these embeddings are passed through dense layers to predict the modality of each sample. By reversing the gradients corresponding to these predictions during back-propagation, we help the feature extractor produce modality-invariant features.

\subsection{Objective Function}

Our goal is to learn a common space where the samples from the same semantic category should be similar, even though they come from different modalities. To learn discriminative features we want to minimize the discrimination loss in both the label space and the representation space, by reducing the cross-modal discrepancy. With regard to the label space, we use a linear classifier to predict the emotion labels of the samples projected in the common space. Outputs of each modality are passed through a sigmoid activation and a binary cross-entropy (BCE) loss $\ell$ is computed. For the cross-modal task we apply a weighted linear combination of those losses: $\mathcal{J}_1 = \lambda_{11} \ell_a + \lambda_{12} \ell_b$. To reduce the cross-modal discrepancy between EEG and music representations, we also compute the BCE loss of the modality prediction after GRL: $\mathcal{J}_2 = \ell_{dd}$. By combining terms $\mathcal{J}_1, \mathcal{J}_2$ we obtain the proposed objective, in which the hyper-parameters $\lambda_i$ control the contribution of each separate component and are determined through trial and error: $\mathcal{J} = \lambda_1\mathcal{J}_1 + \lambda_2\mathcal{J}_2$.

\section{Experimental Setup}

\subsection{The DEAP Dataset}

DEAP \cite{koelstra} is a comprehensive dataset that includes EEG signals of music listening, collected from 32 subjects. Each subject watches forty 1-min long music videos while having their EEG recorded. After each video trial, the subject was instructed to rate the emotion that was elicited upon the entire trial in 5 dimensions: valence, arousal, dominance, liking and familiarity to the track. In this paper we solely experiment with the 2D emotion space, determined by valence and arousal, whose ratings range from 1 (weakest) to 9 (strongest). We use the EEG signals in their already preprocessed form: recorded at a sampling rate of 512 Hz and denoised by bandpass filtering, after downsampling to 128 Hz. Eye-related artefacts were removed whereas the 10-20 electrode placement system was followed.

\textbf{Specifying Music Tracks:} The 40 one-minute music stimuli of DEAP are not included in the dataset, so we located the video clips of the corresponding tracks and isolated the minute of interest for each one, according to the metadata provided. The task of deriving the common latent space poses a crucial challenge: the semantic gap between the ``subjective" affective responses of participants and the emotion tags of the songs. Ideally, we need musical stimuli that are tagged in accordance with the participants' annotations. DEAP stimuli have been selected for this purpose and have been independently annotated by the experimenters at track level. Nearly all songs received average ratings from the participants that were in accordance with those annotations. We found that only $6/40$ songs had such an inconsistency and discarded them. The resulting track set is used to extract MusiCNN embeddings which we make available\footnote{\href{https://github.com/klean2050/EEG\_CrossModal}{https://github.com/klean2050/EEG\_CrossModal}} as well.

\subsection{Input Feature Extraction}

EEG and music signals are processed differently in order to produce an embedding form suitable for multimodal training. DEAP signals are first cut to 3-second segments, while having their preparatory phase removed. For feature extraction purposes we consider differential entropy features, reported to achieve superior performance in the task \cite{critical}. Differential entropy (DE) $h$ is defined as:
\begin{equation}
    h(X)=-\int_{X} f(x) \log (f(x)) d x,
\end{equation}
where $X$ is an EEG segment and $f(x)$ its distribution. Assuming further that the utilized signals can be modeled as Gaussian distributions, i.e. $f(x) = N(\mu, \sigma)$, then $h(X)$ can be determined by the logarithm energy spectrum of $X$ as follows:
\begin{equation}
    \begin{gathered}
h(X)=-\int_{-\infty}^{\infty} \frac{1}{\sqrt{2 \pi \sigma^{2}}} \exp \frac{(x-\mu)^{2}}{2 \sigma^{2}} \\ \log \left(\frac{1}{\sqrt{2 \pi \sigma^{2}}}
\exp \frac{(x-\mu)^{2}}{2 \sigma^{2}} \right) d x=\frac{1}{2} \log 2 \pi e \sigma^{2}.
\end{gathered}
\end{equation}
Thus, for each EEG segment we use the Short-Time Fourier Transform with an 1-sec non-overlapping Hanning window to compute the variance $\sigma^2$ for each of the three windows in the frequency domain and subsequently we compute $h$ in each channel of the four available EEG rhythms: $\theta$ (4-7Hz), $\alpha$ (7-13Hz), $\beta$ (13-30Hz) and $\gamma$ (31-50Hz). The features for all four bands are concatenated and the resulting feature vector is then used as channel-wise input. On the other hand, music tracks are cut in segments aligned with the EEG throughout their whole (3 sec) duration and fed directly into the pre-trained model, from which we extract its ``pool5" embeddings.

\subsection{Evaluation Protocol}

We evaluate our proposed method using accuracy to assess the supervised predictions for each modality and the Precision@10 (P@10) and mean Average Precision (mAP) metrics for the retrieval of music tracks given EEG queries. Those two metrics have been widely used to assess retrieval tasks in the literature \cite{won2020multimodal,dscmr} as they evaluate the response's distance-based ranking to each query. In particular, P@10 considers the top 10 ranked tracks whereas mAP evaluates the whole ranking. Results are also presented after trial aggregation: For the accuracy, we simply denote a prediction as correct by majority voting on the segment-wise predictions. For the retrieval metrics, since no such voting can be made, we consider the median of the segment-wise distance scores as the overall query score. 
\vspace{-0.2cm}

\section{Experiments} \label{sec:exp}

The training procedure considers personalized models, each one trained on data of a specific subject and the respective audio stimuli. To compensate for possible annotation noise, we binarize the emotion labels by setting the threshold to the median score 5, as in \cite{koelstra}. Following the same paradigm, we consider separate experiments for valence and arousal dimensions. We apply 5-fold stratified cross-validation to train each network, where each fold holds 20\% of the total trials (7 tracks). Additionally, we apply class weights to alleviate any subject-specific data imbalance. All networks are optimized using Adam at a $10^{-4}$ learning rate and patience of 15 epochs of non-decreasing validation loss. \vspace{-0.3cm}

\subsection{Predicting Emotion Tags}

We first evaluate the models' performance on Emotion Recognition for both EEG and Music modalities (Table~1). EEG scores show high variance per subject, on average reaching up to $70.4\%$ on valence and $68.9\%$ on arousal after trial aggregation. The obtained scores are competitive for the specific dataset, despite the simple utilized architecture, something we attribute to the impact of music co-training and the adaptation of the common latent space. This contribution is further quantified in Section~5.3. Additionally, aggregating predictions on a per-track basis provides substantially enhanced results compared to non-aggregated ones, with the EEG accuracy increasing by above 5\% in arousal recognition and about 8\% in valence, implying that there is strong correlation (e.g., in the form of clusters) between same-track samples, especially in valence. On the other hand, despite the small number of tracks in our music set, the recognition performance is substantially high for the music branch, 78.8\% average on valence and 91.9\% on arousal, something that indicates the robustness of our transfer learning module. \vspace{-0.3cm}

\begin{table}[!t] \label{tab:pre}
\centering
\begin{tabular}{c|c|c}
\textbf{Dimension} & \textbf{Non-Aggregated} & \textbf{Aggregated} \\ \hline

Valence
         & 62.9\% -- 71.5\%
         & \textbf{70.4}\% -- \textbf{78.7}\% \\
Arousal
         & 63.3\% -- 88.0\%
         & \textbf{68.9}\% -- \textbf{91.9}\%
         
\end{tabular} \vspace{-0.2cm}
\caption{Emotion Accuracy Scores for (EEG -- Music) modalities, reporting mean values over 32 subject-specific models.} \vspace{-0.2cm}
\end{table}

\begin{table}[!t]
\centering
\begin{tabular}{c|c|c}
\textbf{Dimension} & \textbf{Precision@10} & \textbf{mAvg. Precision}\\ \hline

Valence & \textbf{19.4}\% -- \textbf{63.8}\%
         & 18.8\% -- 59.1\%\\
Arousal & 18.4\% -- 65.0\%
         & \textbf{19.9}\% -- \textbf{67.8}\%\\
         
\end{tabular} \vspace{-0.2cm}
\caption{(Track -- Emotion) Retrieval Scores on EEG input queries, reporting mean aggregated scores over 32 subjects.}
\vspace{-0.5cm}
\end{table}

\begin{figure*}
 \centerline{
 \includegraphics[scale=0.33]{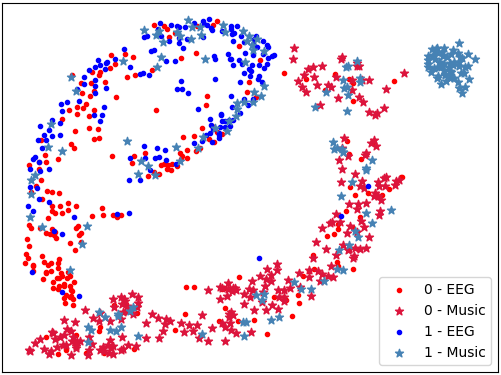}
 \includegraphics[scale=0.33]{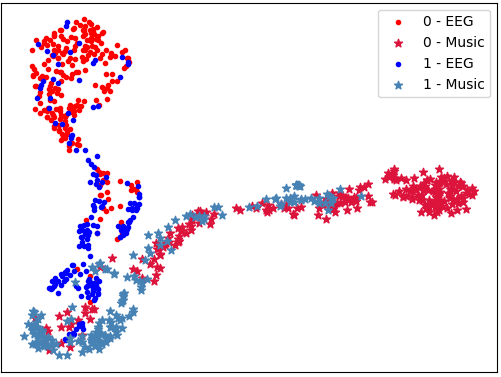}
 \includegraphics[scale=0.33]{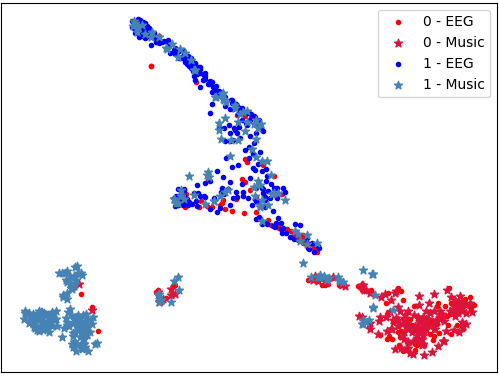}
 \includegraphics[scale=0.33]{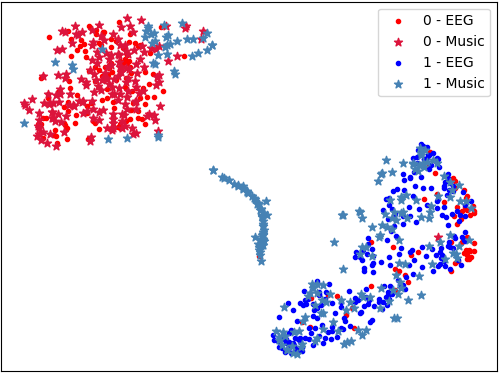}}
 \vspace{-0.25cm}
 \caption{t-SNE visualisation of the common space for subjects --from left to right-- 8, 15 (Valence) and 18, 20 (Arousal). 0 $\rightarrow \text{Low} \,|\, 1\rightarrow$ High}
 \label{fig:tsne}
 \vspace{-0.4cm}
\end{figure*}

\subsection{Retrieving Tracks from EEG Queries}

Table 2 summarizes the retrieval scores from the personalized models, acquired by querying the common representation space of each network with a test EEG sample and then evaluating the ranking of music samples based on their distance to the query. Retrieval metrics provide robust results in both cases, indicating that the EEG samples are well-situated in the common space and the majority of them are capable of retrieving tracks that are emotionally consistent. Specifically, in the case of induced valence, a P@10 value of 63.8\% is achieved. We note that this percentage is higher than the reported mAP (59.1\%), strongly implying that the learned valence space is fragmented into local subspaces of high similarity. Arousal on the other hand seems to be more consistently represented, as both mAP and P@10 median retrieval scores indicate that the majority of tested tracks can derive consistent music rankings, in contrast to valence where the emotional response similarity seems concentrated to the top-ranked elements. As a result, the correct retrieval percentage conditioned on arousal approaches 68\% on average across subjects. We also note some preliminary results in approaching retrieval of the exact stimulus of an EEG sample. The derived scores, around 20\%, are clearly above random selection, however we believe that further experimentation is required on the temporal resolution of the input samples, yielding an interesting direction of future study. \vspace{-0.3cm}

\subsection{Ablation Study}

In our study we incorporated a complex objective function, combining 3 BCE terms to minimize the discrimination loss in both the label space and the common latent space. To further investigate the impact of our proposals on the models' performance, we trained separate sessions, first by considering sole EEG samples without music supervision, and second by avoiding the domain discrimination module. From the results in Table~3 we deduce that our full objective $\mathcal{J}$ leads to higher overall performance, indicating that all utilized terms contribute to richer EEG affective representations. Specifically, we can see that the absence of multimodal training sharply impacts the validity of the common space and reduces the classification performance, 2.6\% in valence and 0.9\% in arousal. On the other side, the absence of domain adaptation causes slighter modifications to the correlation of samples and stimuli, as measured by precision metrics. Through this module we manage though to better distribute samples in the common space, break modality-specific clusters and reduce the overall sample distances (Section 6.1), which is reflected in the improved classification performance in both experiments. \vspace{-0.2cm}

\begin{table}[!t]
\centering
\begin{tabular}{c|c|c|c}
\textbf{Metric} & $\mathcal{J}$ & $\ell_a$ only & $\neg\, \ell_{dd}$\\ \hline
Acc$_{\text{EEG}}$  & \textbf{70.4}\% -- \textbf{68.9}\% & 67.8\% -- 68.0\%
         & 67.9\% -- 63.4\% \\
P@10  & \textbf{63.8}\% -- 65.0\% & 57.3\% -- 53.1\%
         & 63.4\% -- \textbf{66.7}\% \\
mAP  & 59.1\% -- 67.8\% & 51.9\% -- 55.8\%
         & \textbf{59.8}\% -- \textbf{68.1}\% \\
\end{tabular} \vspace{-0.2cm}
\caption{Ablation on the Objective Function for (Valence -- Arousal). Here we solely consider mean aggregated scores over 32 subjects.}
\vspace{-0.5cm}
\end{table}

\begin{figure} \hspace{-0.2cm}
\centerline{
 \includegraphics[scale=0.44]{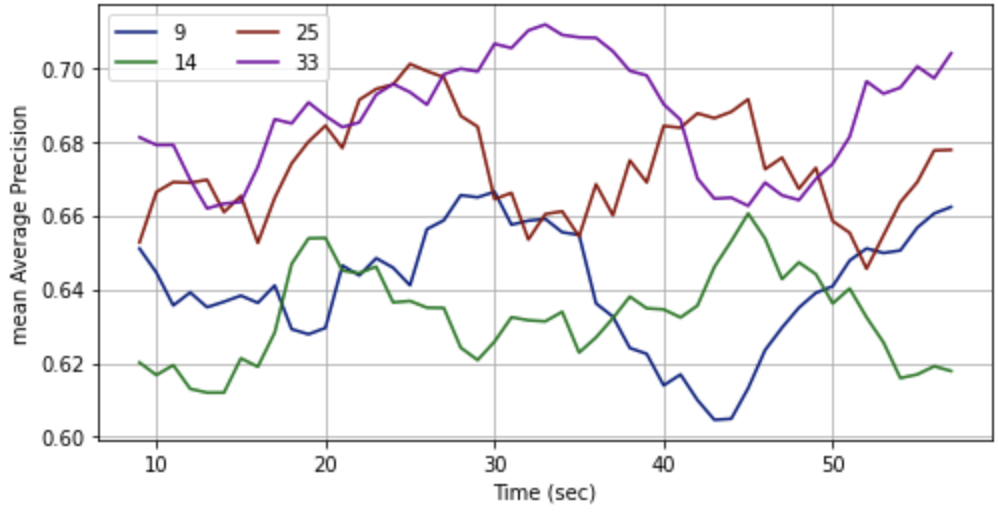}}
 \vspace{-0.4cm}
 \caption{Arousal mAP scores over the 58 time samples for the numbered tracks, averaged across all subjects.}
 \label{fig:temporal}
 \vspace{-0.55cm}
\end{figure}

\section{Qualitative Analysis}

\textbf{Studying the Common Space:} We visually inspect the produced latent space using t-SNE to reduce its 64D dimension to 2D. We select one of the 5 trained models for 2 subjects in Valence and Arousal to display their results in Fig.~\ref{fig:tsne} (similar trends are observed for most subjects). It is evident that latent domain adaptation has alleviated the cross-modal discrepancy and the modalities homogenize their embeddings to a certain degree. Cohesive sub-clusters are though visible, especially in the case of valence. This provides an explanation towards the discrepancy we observed between P@10 and mAP metrics, since the top-ranked track retrievals originate from the corresponding local subspace, but there is no coarse bisection between high- and low- valence samples, in contrast to the case of arousal.

\textbf{Temporal Variation of Recognition:} Since each track is segmented into 58 overlapping samples of 3 sec, it is expected that the emotion is not elicited at the same pace throughout its duration. Hence, the temporal variation of our scores could indicate important moments in the track. In Fig.~\ref{fig:temporal}, we present the temporal evolution of the mAP scores for selected music tracks, averaged across all subjects. While the raw plots are noisy, each song individually exhibits a pattern of variation, which we depict by applying a 7-sample moving average filter. Scores typically reveal an oscillating pattern on the time axis and emotions are highly induced at certain peaks of the graph. These patterns reveal a characteristic picture of emotional induction in songs and could be subject of further experimentation. \vspace{-0.2cm}

\section{Conclusion} \vspace{-0.2cm}
\label{sec:conc}

In this paper we presented a novel approach to analyze emotion induction from EEG recordings of music listening. We proposed a cross-modal framework to learn rich affective representations for EEG data through music supervision and adaptation of a common latent space, from which one could retrieve consistent music rankings from EEG queries. Our approach indicates that distilling information from processed musical stimuli to the respective EEG signals can improve performance and provide insights on personalized emotion analysis. To the best of our knowledge, this is the first study to propose a complete framework for the specific task and dataset, thus our results can be viewed as a concrete baseline. This framework can be used to model the EEG-Music relationship by using different condition mechanisms, e.g., musical beat. Another interesting direction would be to explore improvements in exact stimulus retrieval.

\vfill\pagebreak
\bibliographystyle{IEEEbib}
\bibliography{refs}

\end{document}